\documentclass[twocolumn, superscriptaddress,nofootinbib,floatfix, amsfonts]{revtex4}

\usepackage{amsmath}
\usepackage{bm}
\usepackage{graphicx}
\usepackage{mathrsfs}
\usepackage{footmisc}
\usepackage{multirow}
\usepackage{graphicx}
\usepackage{booktabs, ctable}

\begin{document}

\title{Dilepton photoproduction measures the fluctuations of initial electromagnetic fields in nuclear collisions}

\author{Guansong Li}
\affiliation{Department of Physics, Tianjin University, Tianjin 300350, China}
\author{Kai Zhou}
\affiliation{Frankfurt Institute for Advanced Studies, Ruth-Moufang-Straße 1, D-60438 Frankfurt am Main​, Germany}
\author{Baoyi Chen}
\thanks{Email: baoyi.chen@tju.edu.cn}
\affiliation{Department of Physics, Tianjin University, Tianjin 300350, China}
\affiliation{Institut f\"ur Theoretische Physik, Goethe-Universit\"at Frankfurt,
Max-von-Laue-Str. 1, D-60438 Frankfurt am Main, Germany}

\date{\today}

\begin{abstract}
Dilepton production from two photon interactions $\gamma\gamma\rightarrow l^+l^-$ are 
studied in semi-central and peripheral nuclear collisions.  
Based on Weizs\"acker-Williams approach, It is shown that 
the dilepton photoproduction is proportional 
to the electromagnetic (EM) fields $\sim E^2B^2$  
and therefore sensitive to the magnitude and lifetime of initial EM 
fields which last only for a short time and are hard to be measured in experiments directly. 
We propose dilepton photoproduction as a probe for 
the nuclear charge fluctuations, which are 
crucial for the electric/magnetic field induced chiral and charge particle evolutions. 
We calculate the relative standard deviation of dilepton mass spectrum with event-by-event 
fluctuating nuclear charge 
distributions (and EM fields).

\end{abstract}
\maketitle


In relativistic heavy ion collisions, 
one of the main goals is to study the properties of the deconfined matter, called 
``Quark-Gluon Plasma" (QGP) produced in the hadronic collisions of 
two nuclei~\cite{Gyulassy:2004zy,Shuryak:2004cy,Song:2010mg}. 
In QGP, huge number of light partons are excited and increase the energy 
density of hot medium in the colliding area. 
They expand outward violently due to the large 
spatial gradient of the pressure~\cite{Song:2008si}. 
The evolutions of 
light partons with electric 
charge and chirality are believed to be controlled by the strong interactions. 
In another aspect, nuclei with electric charges $Ze$ 
($e$ is the electron charge) 
are accelerated to nearly the speed of light, and generate 
extremely strong electromagnetic (EM) 
fields with a short lifetime $\sim 2R_A/\gamma_L$~\cite{Deng:2012pc,Tuchin:2013ie}, 
where $R_A$ and $\gamma_L$ are the nuclear radius and the Lorentz factor of fast moving 
nucleons. Besides nuclear hadronic collisions, these 
EM fields can also interact with the target nucleus (moving in the opposite 
direction)~\cite{Klein:1999qj, Shi:2017qep,Adam:2015gba,Khachatryan:2016qhq} 
or the other EM fields 
generated by the target nucleus~\cite{Adams:2004rz,
Baur:2001jj, Baltz:2009jk, Aaboud:2017bwk,Zha:2018ywo,Klein:2018cjh}. 
These reactions 
have been extensively studied in the Ultra-peripheral collisions (UPCs) 
absent of hadronic 
collisions~\cite{Baltz:2007kq,Baur:2007fv, Aaboud:2017bwk, Yu:2015kva,
Klein:2016yzr,Abbas:2013oua}. 
The EM fields can reach its maximum value at the order 
of $eB \sim 10 m_\pi^2$~\cite{Deng:2012pc}
in the semi-central and 
peripheral collisions with the impact parameter around $b\sim 10$ fm at the 
colliding energies of Relativistic Heavy Ion Collider (RHIC) and Large Hadron Collider (LHC).

In the semi-central collisions with the existence of both deconfined medium and 
strong EM fields, the EM fields 
can affect the evolutions of 
charge and chiral partons in the early stage of the 
hot medium expansion, such as 
Chiral-Magnetic-Effect (CME)~\cite{Kharzeev:2007jp} and 
Electric-Seperation-Effect (CSE)~\cite{Huang:2013iia}. 
However, with the background of 
strong collective expansions driven by the 
pressure gradient, 
electric/magnetic field induced parton evolutions 
are contaminated and 
difficult to be quantified with the final hadron spectra~\cite{Shi:2017cpu, Xu:2017qfs}. 
The signals of these effects 
in heavy ion collisions 
are still under debate. Whether these effects are 
observable or not depends sensitively on the magnitude 
and lifetime of the initial electromagnetic fields. 
These EM fields last for a very 
short time and seems impossible to be measured directly. 
In this article, we 
propose that the dilepton photoproduction which is proportional to 
the $\propto B^4$ (or $E^4$) can reveal properties of EM fields 
in the early stage of nuclear collisions~\cite{Shi:2017qep}.

Initial electromagnetic fields can affect chiral/charged particle evolutions, 
and also produce vector mesons ($J/\psi$, $\phi$, \emph{et al}) and dileptons ($e^+e^-$, 
$\mu^+\mu^-$), which have been widely studied in UPCs and is in good agreement 
with the lowest order Quantum Electrodynamics calculations. 
In semi-central nuclear collisions of 
RHIC and LHC energies, 
experiments have observed the 
significant yield enhancement of dileptons at the invariant mass of $J/\psi$, with 
the feature only in extremely low transverse momentum $p_T<0.3$ GeV/c~\cite{Adam:2015gba}. 
This enhancement is far above the hadronic contributions, and is attributed to the 
coherent photon-nuclear interactions: EM fields are approximated as quasi-real photons 
(Weizs\"acker-Williams Method)~\cite{WZ-method, WZ-method-2} which 
scatter with the target nucleus moving in the opposite direction and fluctuate into vector 
mesons. 

At the RHIC Au-Au and U-U collisions, experiments also observe a 
continuum enhancement of $e^+e^-$ in the low invariant mass spectrum 
$0.4\ \mathrm{GeV}<M_{ll}<2.6$ GeV with the limitation of $p_T<0.15$ GeV/c at the impact 
parameter $b\sim 10$ fm in their preliminary results~\cite{Brandenburg:2017meb}.
The STAR continuum observables are compatible with the two-photon production contribution. 
This indicates that dilepton photoproduction dominates the yield in the low 
invariant mass region even in hadronic collisions~\cite{Zha:2018ywo}. 
In this work, we study the dilepton photoproduction with the fluctuating nuclear 
charge distributions, which is considered as an important input 
for the particle evolutions induced by EM fields.

With large Lorentz factor $\gamma_L\sim \sqrt{s_{NN}}/(2m_N)$ where 
$m_N$ and $\sqrt{s_{NN}}$
are the nucleon mass and the colliding energy, the 
transverse electric fields of the fast moving nucleus 
are enhanced by the Lorentz factor, $E_T^{i}=\gamma_L E_T^{i-RF}$. And their magnitude 
is similar with the magnetic fields, 
$|E_T^{i}|\approx |B_T^{i}|$ ($i=$ nucleus 1 or 2).
Here $E_T^{i-RF}$ is the 
electric fields in the nuclear rest frame. Meanwhile, the longitudinal component 
is correspondingly suppressed and therefore negligible.   
These transverse electromagnetic fields can be approximated to be 
a swarm of quasi-real photons moving longitudinally~\cite{krauss:1997}. 
The configuration of EM fields depends on the 
form factor, which is the Fourier transform of the nuclear charge density, 
\begin{align}
F({\bf q}) =\int \rho_{\rm Au}({\bf r}) \exp(i{\bf q}\cdot {\bf r} )d{\bf r}
\end{align}
where $\rho_{\rm Au}({\bf r})$ is the normalized nucleon distribution of Au, 
$\int d{\bf r} \rho_{\rm Au}({\bf r}) =1$. 
The photon spectrum is  
determined by the conservation of  
energy flux through the transverse plane, 
$\int dt d{\bf x}_T  |{\bf E}_T\times {\bf B}_T| = \int dw d{\bf x}_T w n(w,{\bf x}_T)$, 
and then the photon density is  
\begin{align}
\label{eq-photon}
{dN_\gamma\over dwd{\bf x}_T} = {Z^2\alpha\over 4\pi^3 w} |\int_0^{\infty} 
dk_T k_T^2 {F({\bf k}_T^2 +({w\over \gamma_L})^2)\over {\bf k}_T^2 +({w\over \gamma_L})^2} 
J_1(x_Tk_T)|^2
\end{align}
with $\alpha= e^2/(\hbar c)= (4\pi)/137$. $w$ and $k_T$ is the photon energy 
and transverse momentum respectively. $J_1$ is the first kind Bessel function.  
In the collisions with $b<2R_A$, protons in the overlap area are suddenly decelerated 
by hadronic collisions, and do not contribute to the dilepton photoproduction, 
see the schematic diagram below~\cite{Baltz:2009jk}. 
In this work, we focus on the effects of nuclear charge fluctuations on dilepton 
photoproduction. As a comparison, the smooth case 
is also studied where the nucleon density is taken to 
be the smooth Woods-Saxon distribution, 
\begin{align}
\label{eq-woods}
\rho_{\rm Au}({\bf r}) = {\rho_0\over 1+ \exp({r-r_0\over a})}
\end{align}
with $\rho_0={1\over A_{\rm Au}} 0.1694$ $\rm fm^{-3}$, $r_0=6.38$ fm and $a=0.535$ fm. 
For the fluctuating distributions of electric charges (or protons) in 
the nucleus, we generate each proton 
position with Eq.(\ref{eq-woods}) by Monte Carlo simulations in each 
colliding event~\cite{Deng:2012pc}. 
The protons in the nucleus are not taken as point charge, instead, 
they are treated with a finite size to avoid the singularities in EM fields and 
the photon spatial densities. Eq.(\ref{eq-woods}) is also employed for charge 
distribution in proton with different parameters,   
\begin{align}
\label{eq-proton}
\rho_p({\bf r}) = {\rho_{p0}\over 1+\exp( {r-r_{p0}\over a_p})}
\end{align} 
where the origin of coordinate is put at the center of the proton. $r_{p0}=1.2$ fm 
and $a_p=a$ describe the size and shape of a proton.  
$\rho_{p0}=0.0458$ $\rm fm^{-3}$ is fixed 
by the normalization of Eq.(\ref{eq-proton}) to be unit.

\begin{figure}[htb]
{\includegraphics[width=0.4\textwidth]{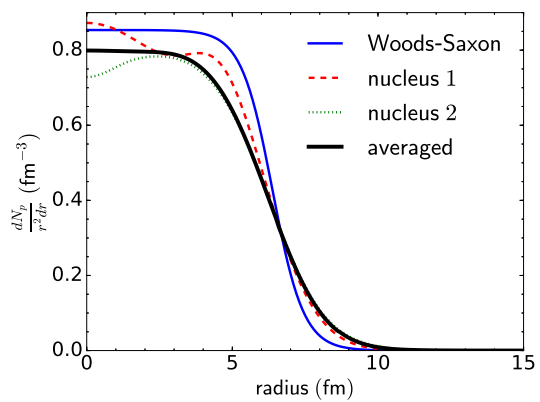}
\caption{ (Color Online) 
Nuclear charge fluctuations in event-by-event MC simulations. The fluctuating 
electric charge densities in \emph{nucleus 1} and \emph{nucleus 2} 
are independent from each other, and deviate from 
the averaged distribution.}
\label{fig-charge} }
\end{figure}

After randomly generating proton positions in the nucleus, 
the fluctuating charge densities in the nucleus can be written as
\begin{align}
\label{eq-averaged}
\rho_{\rm fluct}({\bf r})=\sum_{i=1}^{Z}\rho_p({\bf r}- {\bf r}_i)
\end{align}
where ${\bf r}_i$ is the position of each proton relative to the nuclear center. 
Note that the smooth disposal of proton charges 
slightly extend the nuclear charge distribution to 
the nuclear surface compared 
with Woods-Saxon distribution, cf. black solid line in Fig.\ref{fig-charge}. 
The dilepton yield will be a little different when 
employing the averaged and Woods-Saxon nuclear charge distributions. On the other hand, the 
standard deviation of dilepton photoproduction is less affected, and 
mainly attributed to the fluctuations of charge densities in each colliding events.

\begin{figure}[htb]
{\includegraphics[width=0.35\textwidth]{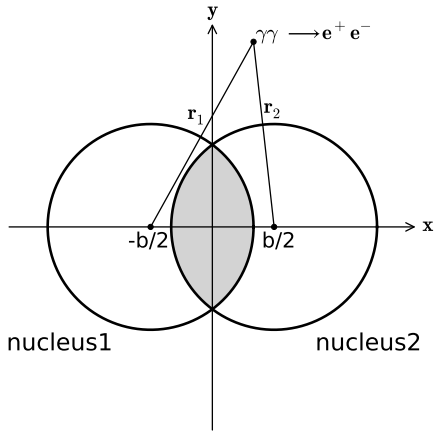}
\caption{
Schematic diagram for two-photon scatterings $\gamma\gamma\rightarrow f\bar f$ 
($f$ is fermion) in the relativistic collisions between nucleus 1 and 2. 
$b$ is the impact parameter. The electric charges in the overlap area do 
not contribute to the two-photon scatterings. 
The origin of coordinates is set in the middle of 
two nuclear centers.  
}
\label{fig-collide}}
\end{figure}

With photon density 
$n(w,{\bf x}_T) \equiv dN_\gamma/(dwdxdy)$ at the transverse coordinate 
${\bf x}_T={\bf r}=(x,y)$, we can calculate the dilepton production from two-photon 
scatterings, $\gamma\gamma\rightarrow l{\bar l}$.  
In this work, we 
neglect the contribution of electric charges 
in the area of hadronic collisions, and 
only consider the 
EM fields generated by spectator protons. 
The EM fields from spectator protons spread 
over the entire transverse plane, which makes quasi-real photons also 
distribute over the entire transverse plane including the 
area inside the nucleus and hadronic collision zone. Therefore, the spatial integration 
of $\gamma\gamma\rightarrow l\bar l$ is over the entire transverse plane, 
see Eq.(\ref{eq-AA-total}) and Fig.\ref{fig-collide}. 
Now we write the dilepton photoproduction in AA collisions with the 
impact parameter $b<2R_A$ as below, 
\begin{align}
\label{eq-AA-total}
{dN\over dM_{l\bar l}} = &\int_{-\infty}^{+\infty} dY\int_0^{2\pi}d\theta 
\int_0^{+\infty} rdr \nonumber \\
&n_1(w_1, r_1) n_2(w_2,r_2) 
 \sigma_{\gamma\gamma\rightarrow l{\bar l}} 
{M_{\gamma\gamma}\over 2}
\end{align}
where $Y={1/2}\ln{(w_1/w_2)}$ and $M_{l\bar l}=2\sqrt{w_1w_2}$ 
are determined by the energies of two scattering photons $w_1$ and $w_2$. 
The distance between the scattering position and two nuclear centers are 
$r_1=\sqrt{b^2/4 +r^2- 2br\cos\theta}$ and $r_2=\sqrt{b^2/4+r^2 +2br\cos\theta}$ 
respectively, and $r$ is the coordinate of the scattering in Fig.\ref{fig-collide}. 
The cross section of two-photon scattering is extracted by the
Breit-Wheeler formula~\cite{Brodsky:1971ud}, 
\begin{align}
\sigma_{\gamma\gamma\rightarrow {l\bar l}} = {4\pi \alpha^2\over M_{l\bar l}^2} 
[&(2+{8m^2\over M_{l\bar l}^2} -{16m^4\over M_{l\bar l}^4}) 
\ln({M_{l\bar l}+\sqrt{M_{l\bar l}^2-4m^2}\over 2m}) \nonumber \\
&- \sqrt{1-{4m^2\over M_{l\bar l}^2}}(1+ {4m^2\over M_{l\bar l}^2})]
\end{align}
where $m$ is the  
lepton mass and $M_{l\bar l}$ is the invariant mass of dilepton.

\begin{figure}[htb]
{\includegraphics[width=0.4\textwidth]{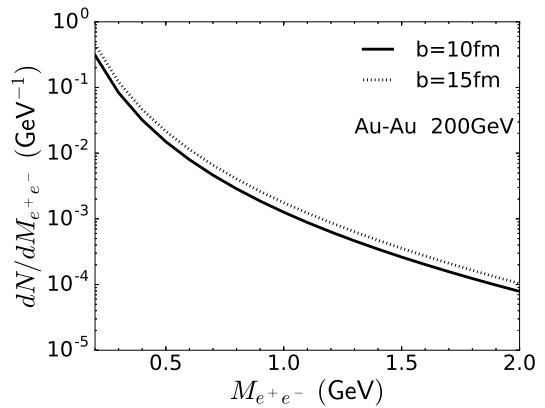}
\caption{
Invariant mass spectra of photoproduced $e^+e^-$ at the impact parameter $b=10$ fm 
(dashed line)
and $b=15$ fm (dotted line) with the Woods-Saxon distribution. 
Note that dilepton production with  
$M_{e^+e^-}>0.4$ GeV is measurable in STAR Collaboration. 
}
\label{fig-dNdM-impact}}
\end{figure}

Before studying the dilepton photoproduction with fluctuating EM fields, we consider the 
situation of smooth charge distribution. 
In Fig.\ref{fig-dNdM-impact}, we calculate the $e^+e^-$ invariant mass spectrum 
$dN/dM_{e^+e^-}$ at the impact parameter $b=10$ fm and $15$ fm 
in $\sqrt{s_{NN}}=200$ GeV Au-Au collisions. In peripheral 
collisions without hadronic collisions, dilepton production only come from 
the interactions of EM fields generated by two nuclei. 
In the situation with $b=10$ fm, final dilepton yields consist of photoproduction from  
two-photon scatterings and thermal emission from QGP. 
In the low invariant mass spectrum, 
dilepton production is dominated by the photoproduction, and we neglect the contribution 
of QGP which is the main source at larger invariant mass region. 
In semi-central collisions, 
we exclude the contribution of electric charges inside the overlap of 
two nuclei~\cite{Baltz:2009jk,Zha:2018ywo}. 
Considering that realistic nuclear charge distribution is continuous, and EM fields inside 
the nucleus is non-zero, the photon spatial density inside the nucleus is also non-zero and 
contributes to the dilepton photoproduction in our calculations. 

\begin{figure}[!htb]
{\includegraphics[width=0.4\textwidth]{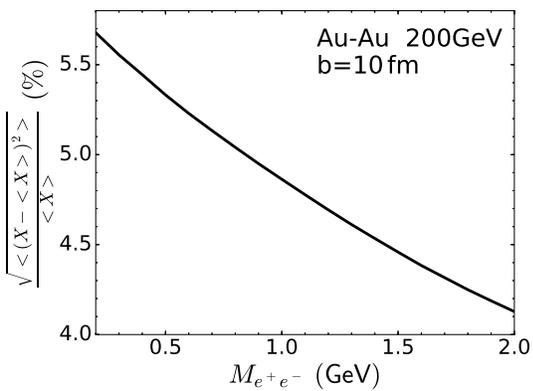}
\caption{
Relative standard deviation 
$\sqrt{\langle (X-\langle X\rangle )^2\rangle}\over \langle X\rangle$
for dilepton photoproduction with event-by-event fluctuating EM fields as a function of 
invariant mass. 
$X\equiv dN/dM_{e^+e^-}$ is the dilepton spectrum.
}
\label{fig-devia}}
\end{figure}

Now, we take the fluctuating charge distribution $\rho_{\rm fluct}({\bf r})$ 
to calculate the photon densities and the dilepton photoproduction. 
The fluctuations of proton positions in the transverse plane is random, which 
make nuclear charge distribution non-isotropic anymore. 
In order to simplify the numerical calculations of photon density with Eq.(\ref{eq-photon}), 
we only consider the radial fluctuations in $\rho_{\rm fluct}$ and employ the 
fluctuating isotropic nuclear charge density (cf. Fig.\ref{fig-charge}) 
in the following calculations. 
With different charge distributions in each event of Au-Au collisions, 
the strength of EM fields is different. If the protons are distributed in the area of 
hadronic collisions due to the fluctuations, they will not contribute to the 
dilepton photoproduction, and $Z_{\rm eff}$ of this event 
will be smaller than its mean value. 
We can obtain the fluctuations of dilepton production induced by the nuclear 
charge fluctuations. 
Considering that particle yield is a scalar observable, the average of 
dilepton yields over many events will partially eliminate the effects of fluctuations. 
Therefore, we evaluate the relative standard deviation of dilepton production, 
$\sqrt{\langle (X-\langle X\rangle )^2\rangle}\over \langle X\rangle$. Here 
$X \equiv dN/dM_{e^+e^-}$ is the dilepton production in one configuration of 
fluctuating EM fields, and 
$\langle X\rangle $ is the averaged yield over many events. 
This observable is weakly affected by   
the uncertainties of two-photon scattering cross sections $\sigma_{\gamma\gamma\rightarrow 
l\bar l}$ and is mainly attributed to the fluctuations of photon densities (EM fields). 
The ratio is around $(4\sim 5)\%$ 
in the experimentally measurable region of invariant mass 
$0.4<M_{e^+e^-}< 2$ GeV, cf. Fig.\ref{fig-devia}. 

In summary, we calculate the dilepton photoproduction from electromagnetic fields in 
peripheral and semi-central collisions. In the low invariant mass region and 
close-to-peripheral collisions, dilepton yields from hot medium radiation produced 
in the nuclear hadronic collisions are relatively small compared with the photoproduction. 
We simulate the fluctuations of proton positions in the nucleus, which results in 
event-by-event fluctuating electromagnetic fields. We calculate the dilepton photoproduction 
based on fluctuating EM fields, and show the relative standard deviation 
$\sqrt{\langle (X-\langle X\rangle )^2\rangle}\over \langle X\rangle$
of their yields $X\equiv dN/dM_{l\bar l}$. 
It is less affected by the dilepton cross section 
$\gamma\gamma\rightarrow l\bar l$, and mainly 
depends on the spatial configurations of quasi-real photons (or EM fields), which 
can help revealing the fluctuations of initial EM fields in the nuclear collisions. 
In the next step, 
we will focus on the polarization of photoproduced particles, which 
is more sensitive to the combined EM fields of two nuclei and their fluctuations.

\vspace{0.4cm}
\appendix {\bf  Acknowledgement}: 
BC acknowledges helpful discussions with Enrico Speranza. 
This work is supported by NSFC Grant No. 11705125, 11547043 and 
Sino-Germany (CSC-DAAD) Postdoc Scholarship.
KZ kindly acknowledge support by the AI grant of SAMSON AG, Frankfurt.

\end{document}